\documentclass{aa}
\usepackage{txfonts}
\usepackage{graphicx}   
\usepackage{amssymb}    
\usepackage{epstopdf}
\usepackage[font=footnotesize,labelfont=bf]{caption}
\usepackage{color}
\usepackage{amsmath}
\usepackage{mathtools}
\usepackage{booktabs,caption}
\usepackage[flushleft]{threeparttable}
\usepackage{hyperref}
\usepackage{csquotes}
\usepackage{enumitem}
\usepackage{orcidlink}
\setlist[itemize]{label=$\bullet$} 
\usepackage{perpage}
\def\apjs{{\rm ApJS}}                  
\def\apjs{{\rm ApJ}}                   
\def\apjl{{\rm ApJ Let}}

\def\a4{\hsize 17.0cm \vsize 25.cm}

\begin{document}

   \title{Dusty clump survival in supernova ejecta}

   \subtitle{Dust-mediated growth vs. crushing by the reverse shock}

\author{
  Sergio Mart\'{\i}nez-Gonz\'alez%
  \thanks{Corresponding author. E-mail: sergiomtz@inaoep.mx}%
  \inst{1}\orcidlink{0000-0002-4371-3823}
}

\institute{
  Instituto Nacional de Astrof\'\i sica, \'Optica y Electr\'onica, AP 51, 72000 Puebla, M\'exico
}
             
   \date{Received: July 13th, 2025; Accepted: September 7th, 2025}
    \authorrunning{Martínez-González}

\abstract
  {Understanding the interaction of dense, cold ejecta clumps with a fast reverse shock, an instance of the \enquote{cloud-crushing} problem, is essential to assess whether core-collapse supernovae act as net dust factories or net dust destroyers.}
  {This work assesses whether dusty ejecta clumps are destroyed by the reverse shock or instead cool, condense, and grow in mass under realistic supernova‑remnant conditions.}
  {Cloud‑crushing timescales are computed and compared to radiative cooling timescales, including both gas‑phase cooling and dust-induced cooling, for a large grid of clump densities, dust‑to‑gas mass ratios, and shock velocities.}
  {When the dust-to-gas mass ratio exceeds $10^{-3}$, gas–grain collisions become efficient enough that the cooling timescale $t_{\rm cool}$ falls below the cloud‑crushing timescale $t_{\rm cc}$ over a broad span of clump densities and shock velocities, enabling dusty clumps to survive even fast reverse shocks. For example, at clump densities $\gtrsim 2 \times 10^{4}$ cm$^{-3}$, dust-to-gas mass ratios $\sim 10^{-2}$, and shock velocities up to $2000$ km s$^{-1}$, enhanced gas–grain cooling drives the system into a regime where dusty clumps can gain additional cold mass and increase their dust masses.}
  {Strong radiative cooling can shield dust‑rich clumps in supernova remnants, enabling a significant fraction of ejecta dust to be injected into the interstellar medium. These results mirror the \enquote{growth} regime found in studies of circumgalactic clouds and rapidly cooling shocked stellar winds, implying a larger dust survival in supernova remnants. Indeed, the dusty globules seen in the Crab Nebula occupy the predicted survival regime across a wide range of physical parameters.}


   \maketitle

\section{Introduction}

The interaction of a dense, cold clump of gas with a hot shocked flow (the \enquote{cloud-crushing problem}) has been extensively studied in the context of supernova remnants (SNRs), where dense clumps of ejecta (and any dust they contain) can be hit by the reverse shock. Hydrodynamic simulations by \citet{Silviaetal2010,Silviaetal2012} examined dust sputtering in idealized clumpy ejecta. They found that small grains ($a\lesssim0.1$$\mu$m) are quickly sputtered and destroyed, whereas larger grains survive the shock interaction. They also showed that higher shock velocities lead to up to $\sim50\%$ more destruction, and that metal-rich ejecta can enhance cooling and modify the destruction efficiency. More recent numerical studies of dust in clumpy supernova ejecta \citep{Kirchschlageretal2019,Kirchschlageretal2023,Kirchschlager2024a} have confirmed these trends. For example, \citet{Kirchschlageretal2019} used hydrodynamic simulations of Cas A knots and found that up to $40\%$ of the dust mass can survive the reverse shock, especially for intermediate grain sizes or high-density clumps. They also found that small grains and certain compositions are more easily destroyed, and that grain--grain collisions can alter the size distribution of the survivors. Including magnetic fields, \citet{Kirchschlageretal2023} showed that field orientation and clump density contrast strongly affect survival: for perpendicular fields, small grains are completely destroyed while large grains ($\sim1$$\mu$m) survive with high efficiency; for higher density contrast, even nanometer grains can survive in part. Finally, \citet{Kirchschlager2024a} studied a time-dependent model of Cas A and found that most dust is processed in the first few hundred years, with survival fractions ranging from $\sim17\%$ for 1 nm grains to $\sim28\%$ for 1000 nm grains integrated over the remnant age. One can note that dust-induced cooling has not been incorporated into such models.

Dust in SNRs may also be observed directly in surviving clumps. For instance, \citet{Grenmanetal2017} catalogued small dusty globules in the Crab Nebula (the remnant of SN 1054). These globules are $\sim400-2000$ AU in size (peaking at $\sim500$ AU), contain $10^{-6}$--$10^{-5} $M$_\odot$ of dust each, and appear as dark spots against the synchrotron background. Their spatial distribution and alignment with optical filaments suggest that they are dense knots of ejecta that have retained their dust. The dust extinction law in these globules matches normal ISM dust, implying survival of typical grains. These observations motivate models of clumpy SN ejecta in which dust can indeed endure shock passage. Thus, understanding the interplay of cooling, shock dynamics, and additionally clump density is crucial to predicting how much SN dust is ultimately injected into the ISM.

Analytical treatments of cloud/clump survival have advanced in recent years. \citet{Gronkeetal2018,Gronkeetal2020} \citep[see also][]{Lietal2020} developed a criterion comparing the cooling time of the turbulent mixing layer at the cloud–flow interface to the cloud–crushing time, identifying regimes of destruction versus survival and growth. In the growth regime, rapid cooling makes mixing \enquote{accretive}: freshly mixed gas condenses onto the cold phase, the shear layer thins, and the growth of hydrodynamical instabilities is suppressed. This approach is adopted here and adapted to the dense, dust-enriched clumps observed in SN ejecta, where the characteristic timescale for a shock to crush the clump (the cloud-crushing time) is compared to the radiative cooling time of the shocked gas within and around the clump. In addition, radiative cooling induced by gas–grain thermal collisions is consistently incorporated into this framework \citep[][]{Dwek1987,MartinezGonzalezetal2016}.

The structure of the paper is as follows: Section \ref{sec:models_methods} presents the analytic model and methods; Section \ref{sec:space} explores the parameter‐space of shock velocities, clump densities and dust-to-gas mass ratios; and Section \ref{sec:conclusions} summarizes the main conclusions and discusses their implications.

\section{Analytic model and methods}
\label{sec:models_methods}
\subsection{Cloud‐crushing timescale}
\label{subsec:crushing}

Consider a spherical clump of radius $R_{\rm cl}$ and density $n_{\rm cl}$ embedded in an ejecta flow of density $n_{\rm ej}$ and shock velocity $v_{\rm sh}$.  When the planar shock encounters the clump, it drives a transmitted shock into the clump; the classical cloud‐crushing time is defined as \citep{McKeeOstriker1977,Klein1994}

\begin{equation}
t_{\rm cc} \;\approx\; \frac{\sqrt{\chi} R_{\rm cl}}{v_{\rm sh}} , 
\qquad 
\chi \;\equiv\;\frac{n_{\rm cl}}{n_{\rm ej}} ,
\end{equation}

which represents the time required for the shock to traverse the clump\footnote{Recent hydrodynamical and magnetohydrodynamical simulations 
(e.g. \citet{Kirchschlager2024a} and references therein) 
demonstrate that the effective \enquote{cloud-crushing} time can deviate 
significantly from the classical \citet{McKeeOstriker1977} estimate. The analytic 
$t_{\rm cc}$ used here should be regarded as a benchmark scaling,}. The post‐shock temperature of the clump gas is given by the Rankine–Hugoniot jump conditions for a strong shock \citep{Zeldovich1967pswh},

\begin{equation}
T_{\rm cl,sh} \;\simeq\; \frac{3 \mu v_{\rm sh}^2}{16 k_{B}} , 
\qquad 
\mu \approx 0.6 m_{H}, 
\end{equation}

where $m_{H}$ is the hydrogen mass and $k_{B}$ is the Boltzmann constant. Immediately behind the shock, the clump density is $n_{\rm sh} = 4 n_{\rm ej}$, and the mixed density between shocked clump and ejecta is taken as

\begin{equation}
n_{\rm mix} \;=\;\sqrt{n_{\rm cl} n_{\rm sh}} \;=\; 2 \sqrt{n_{\rm cl} n_{\rm ej}} .
\end{equation}

To assess whether the clump cools radiatively before being destroyed, one can compute the ratio between the cooling and the cloud crushing times \citep{Gronkeetal2018,Gronkeetal2020,Lietal2020}

\begin{equation}
\frac{t_{\rm cool}}{t_{\rm cc}} 
\;=\; 
\frac{3 k_{B} T_{\rm cl,sh}}{2n_{\rm mix} \Lambda_{\rm tot}(T_{\rm cl,sh})} 
\;\times\;\frac{v_{\rm sh}}{\sqrt{\chi} R_{\rm cl}}=\frac{L_{\mathrm{cool}}}{\sqrt{\chi} R_{\mathrm{cl}}},
\end{equation}

where $\Lambda_{\rm tot}(T)$ is the total cooling function (gas+dust) evaluated at the post‐shock temperature $T_{\rm cl,sh}$, and $L_{\mathrm{cool}}\equiv v_{\mathrm{sh}} t_{\mathrm{cool}}$ is the post‐shock cooling length. Accordingly, one may define the clump–frame cooling length $\ell_{\mathrm{cool,cl}} \equiv L_{\mathrm{cool}}/\sqrt{\chi}$,
so that the ratio takes the compact form 
\begin{equation}
\frac{t_{\mathrm{cool}}}{t_{\mathrm{cc}}} = \frac{\ell_{\mathrm{cool,cl}}}{R_{\mathrm{cl}}}   .
\end{equation}

Note that since $\chi=n_{\rm cl}/n_{\rm ej}$ enters both $t_{\rm cc}\propto\chi^{1/2}$ and 
$n_{\rm mix}=2n_{\rm cl}\chi^{-1/2}$, $n_{\rm ej}$ 
cancels out of the ratio, leaving $t_{\rm cool}/t_{\rm cc}$ independent of $n_{\rm ej}$. 

\subsection{Radiative cooling times}
\label{subsec:cooling}

\begin{figure}[t]
\centering
\includegraphics[width=\columnwidth]{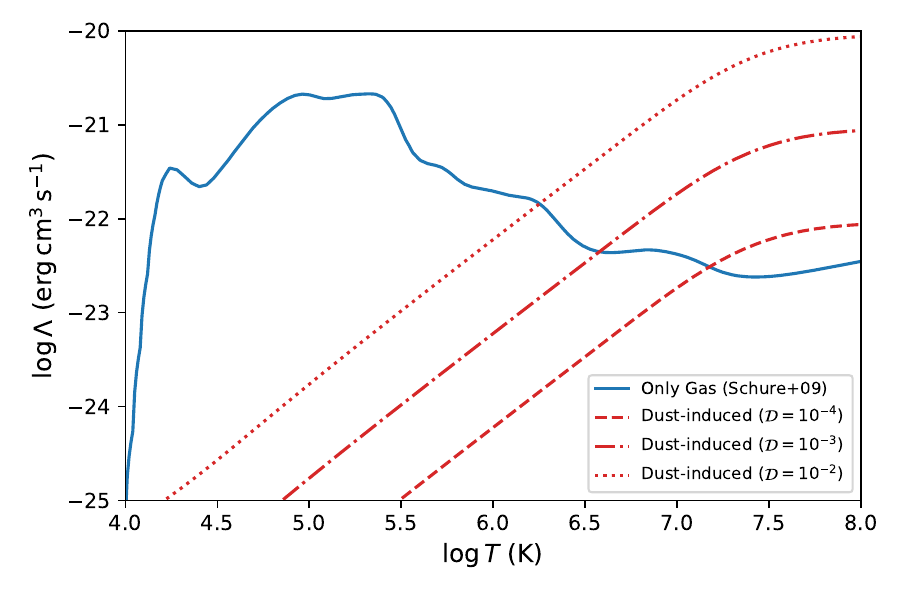}
\caption{Cooling rate per unit volume vs.\ temperature. Solid curve: gas-phase cooling \cite{Schureetal2009}. Dashed, dash-dotted, dotted: dust-induced cooling for $\mathcal{D}=10^{-4}, 10^{-3}, 10^{-2}$, respectively, assuming a log-normal grain-size distribution (5 nm $<a<$ 0.5 $\mu$m).}
\label{fig:cooling}
\end{figure}

The total cooling rate per unit volume at temperature $T$ is written as the sum of gas‐phase cooling and dust‐induced cooling,

\begin{eqnarray}
\Lambda_{\rm tot}(T) &=& \Lambda_{\rm gas}(T) + \Lambda_{\rm dust}(T) ,
\end{eqnarray}

where $\Lambda_{\rm gas}(T)$ is obtained by interpolating the tabulated $\log T$–$\log\Lambda$ data of \citet{Schureetal2009} above $10^4$ K, and \citet{DalgarnoandMcCray1972} for $\leq 10^4$ K. The dust‐induced cooling curves, arises from inelastic collisions between thermal electrons (and ions) and dust grains, were calculated as prescribed by \citet{Dwek1987} \citep[see Figure \ref{fig:cooling} and Appendix A in][]{MartinezGonzalezetal2016}. A log-normal grain‐size distribution is assumed \citep{MartinezGonzalezetal2021},

\begin{eqnarray}
\frac{dn}{da} &\propto& \frac{1}{a}
\exp\!\Bigl[-\frac{\ln^{2}(a/a_{\rm peak})}{2\sigma^{2}}\Bigr] ,
\end{eqnarray}

truncated between $a_{\min}=0.005$ $\mu$m and $a_{\max}=0.5$ $\mu$m, with $a_{\rm peak}=0.10$ $\mu$m and $\sigma=0.7$, and normalized to the assumed value of $\mathcal{D}$. 

\begin{figure*}[ht]
\centering
\includegraphics[width=1.8\columnwidth]{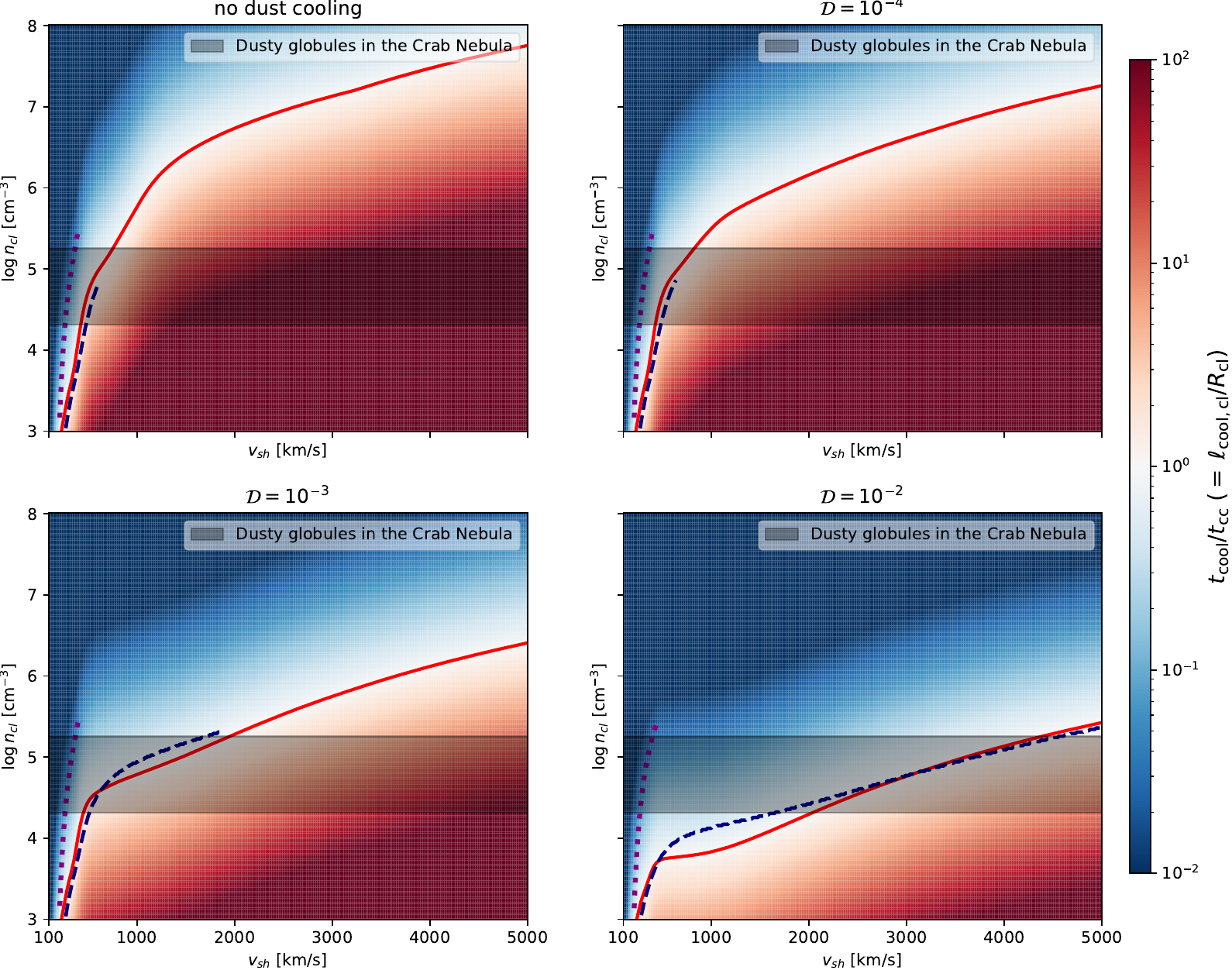}
\caption{Regimes of clump survival versus crushing for clumps with $R_{\rm cl}=500$ AU. The maps show $t_{\rm cool}/t_{\rm cc}$ as a function of shock velocity and clump density. The panels correspond to cases with metal-line cooling only (top left), and different dust-to-gas mass ratios: $\mathcal{D}=10^{-4}$ (top right), $\mathcal{D}=10^{-3}$ (bottom left), and $\mathcal{D}=10^{-2}$ (bottom right). The red solid lines mark $t_{\rm cool}=t_{\rm cc}$. Clumps above the line (lower $v_{\rm sh}$ or higher $n_{\rm cl}$) can cool/grow, while below the line they are crushed. Higher dust-to-gas mass ratios expand the survival region. Blue dashed: isobaric $\mathcal{R}=1$; purple dotted: isochoric $\mathcal{R}=1$. Isochoric crossings occur only at $v_{\rm sh}\!\lesssim$ a few $10^2$ km s$^{-1}$; realistic clumps fall between these limits. In both cases, regions above the $\mathcal{R}=1$ lines correspond to grain growth exceeding thermal sputtering. The hatched area indicates the range of observed dusty globules in the Crab Nebula.}
\label{fig:space}
\end{figure*}

\subsection{Grain processing}
\label{subsec:processing}

Within the shocked clump layer, the gas undergoes cooling from its post-shock temperature $T_{\rm cl,sh}$ until either radiative losses cease or a thermal floor is reached. Grain processing operates only within restricted intervals: 
thermal sputtering requires $T \gtrsim 5\times 10^5$ K \citep{Bocchioetal2014}, 
while grain growth by accretion occurs for $T \lesssim 2\times 10^{3}$ K \citep{Nozawaetal2012}. 
The effective timescales of each regime along the cooling trajectory are

\begin{eqnarray}
t_{\mathrm{sput}} &=& C
\int_{5\times10^5 {\rm K}}^{T_{\rm cl,sh}}
\frac{T}{\Lambda_{\mathrm{tot}}(T)} \mathrm{d}T,\\[0.6em]
t_{\mathrm{grow}} &=& C
\int_{T_{\rm floor}}^{2\times10^{3} {\rm K}}
\frac{T}{\Lambda_{\mathrm{tot}}(T)} \mathrm{d}T ,
\end{eqnarray}

where $C=\tfrac{5}{2}k_B^2/P$ for isobaric cooling and $C=\tfrac{3}{2}k_B/n$ for isochoric cooling, with $P=n(T)k_BT$ and $T_{\rm floor}=100$ K the terminal gas temperature. If the gas does not cool from $10^6$ K to $10^{3}$ K within one $t_{cc}$, then $t_{\rm grow}$ is set to 0 s.

The sizes of dust grains are the result of the balance between the accretion growth rate ($\dot{a}_{\mathrm{grow}}$) and the thermal sputtering rate ($\dot{a}_{\mathrm{sput}}$). Evaluated along the cooling trajectory, their cumulative impact is captured by

\begin{eqnarray}
 \mathcal{R}\;=
\frac{t_{\mathrm{grow}} \dot{a}_{\mathrm{grow}}}{t_{\mathrm{sput}} \left|\dot{a}_{\mathrm{sput}}\right| }   ,
\end{eqnarray}

with $\mathcal{R}>1$ indicating net growth and $\mathcal{R}<1$ net destruction within the shocked region. 

The accretion growth rate, as a function of the gas number density $n$, and temperature $T$, is evaluated as \citep{Spitzer1978}

\begin{eqnarray}
\dot{a}_{\mathrm{grow}} &=& 
\frac{3 S_{\mathrm{stick}} \mu_{\mathrm{mol}} n}
     {4 \rho_{\mathrm{gr}}}
    \sqrt{\frac{8 k_B T}{\pi m_{\mathrm{s}}}}   ,
\end{eqnarray}

with $S_{\mathrm{stick}}=1$ \citep[justified for grain temperatures $\lesssim 30$ K and binding energies $\gtrsim 0.1$ eV ($\approx 1100$ K),][]{Ferraraetal2016}, grain density $\rho_{\mathrm{gr}}=2.26$ g cm$^{-3}$, and key species mass $m_{\mathrm{s}}=12$ $m_{H}$ (carbon). The thermal sputtering rate is calculated using the approximation made by \citet{TsaiMathews1995} to the results of \citet{Tielensetal1994} and \citet{DraineandSalpeter1979},

\begin{eqnarray}
\dot{a}_{\mathrm{sput}} &=& -1.4 n h 
\left[ \left(\frac{10^6 {\rm K}}{T}\right)^{2.5} + 1 \right]\, ,
\end{eqnarray}

where $h=3.2\times10^{-18} {\rm cm^4 s^{-1}}$. Apart from thermal sputtering, other prominent destruction/disruption channels are kinetic sputtering and shattering due to grain--grain collisions \citep[e.g.][]{Kirchschlageretal2019}.

\section{Parameter-space analysis}
\label{sec:space}

The analysis covers a grid across shock velocities $v_{\rm sh}=100$--$5000$ km s$^{-1}$ and clump densities $n_{\rm cl}=10^{3}$--$10^8$ cm$^{-3}$, with the ratio $t_{\rm cool}/t_{\rm cc}$ computed for each case. The dust-to-gas mass ratio $\mathcal{D}$ is varied from 0 to $10^{-2}$ to assess the role of dust-induced cooling. The clump radius is fixed at $R_{\rm cl}=500$ AU (the peak of the size distribution derived by \citet{Grenmanetal2017} for the dusty globules in the Crab Nebula).

Figure \ref{fig:space} shows the boundary at $t_{\mathrm{cool}}/t_{\mathrm{cc}}=1$, where the transmitted shock changes from adiabatic to radiative. For $t_{\rm cool}<t_{\rm cc}$, radiative losses confine significant shock-heating to a superficial layer of depth $\approx \ell_{\mathrm{cool,cl}}$, limiting grain exposure to temperatures favorable for thermal sputtering and potentially leading to preserve or even increase its cold gas mass through condensation from the shocked flow \citep{Gronkeetal2018}. For $t_{\rm cool}>t_{\rm cc}$, the shock remains quasi-adiabatic, leading to clump disruption.

Crucially, as $\mathcal{D}$ grows beyond $\sim10^{-3}$, the cooling timescale falls below the cloud-crushing time across a broader range of clump densities and shock velocities, shifting the survival boundary towards higher shock speeds. For example, at $n_{\rm cl} = 2\times10^4$ cm$^{-3}$, a clump with $\mathcal{D} = 10^{-2}$ can withstand shocks up to $\sim2000$ km s$^{-1}$, whereas in the absence of dust it would be destroyed at velocities as low as $\sim500$ km s$^{-1}$. 

In the radiative regime, the growth of Kelvin–Helmholtz instabilities are expected to be suppressed \citep{Gronkeetal2020}. One could also anticipate that the fast weakening of the transmitted shock may suppress kinetic sputtering, another important channel for grain destruction \citep[][and references therein]{Kirchschlager2024a}; however, this needs to be tested via full magneto–hydrodynamical simulations.

The blue dashed (isobaric) and purple dotted (isochoric) curves mark $\mathcal{R}=1$. In the isobaric case, the curve is most extended for $\mathcal{D}=10^{-2}$ (bottom right panel), but short segments are also present in the other panels. Wherever this curve appears, the shocked gas cools to $\lesssim 2 \times 10^{3} \mathrm{K}$ within one $t_{\mathrm{cc}}$, so grain growth at least balances and locally exceeds thermal sputtering, yielding a net positive effect on dust survival. For $\mathcal{D}=10^{-2}$, the $t_{\mathrm{cool}}/t_{\mathrm{cc}}=1$ and $\mathcal{R}=1$ contours track each other closely in the $n_{\rm cl}$--$v_{\rm sh}$ plane, whereas at lower $\mathcal{D}$ their overlap is confined to a much narrower region. In the isochoric extreme $\mathcal{R}$ exceeds unity only at $v_{\rm sh}\sim100$--$300$ km s$^{-1}$.

The analytic regime diagrams also show that the dusty globules in the Crab Nebula are well within the survival region for a wide range of parameters. \citet{Grenmanetal2017} report typical clump dust mass densities of $\sim8\times10^{-22}$ to $\sim7\times10^{-21}$ g cm$^{-3}$, which, assuming a dust-to-gas mass ratio of $\mathcal{D}=10^{-2}$ and a mean molecular weight of $\mu_{mol}=2.33m_{H}$, translates into $n_{\rm cl}\sim 2\times10^4$–$1.8\times10^5$ cm$^{-3}$. These values lie in the regime where $t_{\rm cool}\ll t_{\rm cc}$, consistent with the observed presence of cold dust.

As shown by \citet{Gronkeetal2020b}, under strong pressure perturbations and high density contrasts, a cloud/clump may undergo fragmentation into fine droplets rather than remaining coherent; the present analytic framework does not account for this potential fragmentation pathway, nor the differing dust survival prospects in droplet ensembles.

\section{Conclusions}
\label{sec:conclusions}

This work adapts the analytic cloud-crushing model of \citet{Gronkeetal2018,Gronkeetal2020} and \citet{Lietal2020} to the case of dusty clumps in SN ejecta. By comparing the cloud-crushing time to the cooling time (including gas and dust cooling), this work shows that many SN clumps lie in a regime where cooling is so efficient that the shock-compressed gas can radiate away its energy, allowing the clump to remain cold and even gain mass. Dust plays a crucial role by boosting radiative cooling. This means that, contrary to the expectation of wholesale destruction, a significant fraction of dusty clumps in SNRs may survive the reverse shock.

This regime is analogous to the \enquote{growth} regime identified in circumgalactic clouds \citep{Lietal2020} and rapidly cooling shocked stellar winds \citep{Palousetal2014,Wunschetal2017}. Our results imply that clumps in e.g. young SNRs like Cas~A or the Crab Nebula can harbor surviving dust. In particular, the dusty globules in the Crab nebula fall in the predicted survival zone, suggesting they are simply clumps whose cooling time was much shorter than their destruction time.

In summary, dust-mediated cooling can protect ejecta clumps from destruction and may even enhance cold-mass growth and dust grain growth in SNRs. This supports the picture that core-collapse supernovae can inject dust into the ISM more effectively than previously thought, as many clumps act as self-shielded reservoirs that deliver dust to the interstellar medium \citep[see also][and references therein]{MartinezGonzalezetal2025}.

Finally, the present analytic framework, built on the cloud crushing paradigm, does not capture magnetic fields, grain--grain collisions, and charge dependent (Coulomb and Lorentz) forces, nor other non linear gas dynamical effects. Therefore the condition $t_{\rm cool}/t_{\rm cc}<1$ should be regarded as an indicator for enhanced chances for dust survival, but not as a sufficient criterion. Dedicated, high resolution magneto-hydrodynamical simulations that include dust microphysics and that compare grain growth and destruction rates are required to calibrate $t_{\rm cc}$ beyond the classical definition by \citet{McKeeOstriker1977} and to quantitatively test the clump survival pathways suggested by \citet{Gronkeetal2020,Gronkeetal2020b} and the reverse shock models of \citet{Kirchschlageretal2019,Kirchschlageretal2023,Kirchschlager2024a} when dust-induced cooling is considered.

\begin{acknowledgements}
The author thank the anonymous referee for their helpful comments and suggestions which improved the quality of the paper.
\end{acknowledgements}

\bibliographystyle{aa}
\bibliography{sample7}

\end{document}